\documentclass[12pt,preprint]{aastex}
\shorttitle{SMA $^{12}$CO(J = 3 -- 2) observations of M51}
\shortauthors{Matsushita et al.}
\slugcomment{Accepted for publication in ApJ Letters}

\begin{document}

\title{SMA $^{12}$CO(J = 3 -- 2) interferometric observations of
	the central region of M51}

\author{Satoki Matsushita\altaffilmark{1},
	Kazushi Sakamoto\altaffilmark{2},
	Cheng-Yu Kuo\altaffilmark{1},
	Pei-Ying Hsieh\altaffilmark{3},
	Dinh-V-Trung\altaffilmark{1},
	Rui-Qing Mao\altaffilmark{4},
	Daisuke Iono\altaffilmark{2,5},
	Alison B. Peck\altaffilmark{2},
	Martina C. Wiedner\altaffilmark{2,6},
	Sheng-Yuan Liu\altaffilmark{1},
	Nagayoshi Ohashi\altaffilmark{1},
	\and
	Jeremy Lim\altaffilmark{1}
	}

\altaffiltext{1}{Academia Sinica Institute of Astronomy and
	Astrophysics, P.O.\ Box 23-141, Taipei 106, Taiwan, R.O.C.;
	satoki@asiaa.sinica.edu.tw}
\altaffiltext{2}{Harvard-Smithsonian Center for Astrophysics,
	60 Garden St., MS-78, Cambridge, MA 02138}
\altaffiltext{3}{Department of Physics,
	National Taiwan Normal University, 88 Sec.\ 4, Ting-Chou Road,
	Taipei 116, Taiwan, R.O.C.}
\altaffiltext{4}{Purple Mountain Observatory,
	Chinese Academy of Sciences, Nanjing 210 008, P.R.\ China}
\altaffiltext{5}{University of Massachusetts,
	Department of Astronomy, Amherst, MA 01003}
\altaffiltext{6}{1.\ Physikalisches Institut, Universit\"at zu K\"oln,
	50937 K\"oln, Germany}

\begin{abstract}
We present the first interferometric $^{12}$CO(J = 3 -- 2)
observations (beam size of $3\farcs9\times1\farcs6$ or
160 pc $\times$ 65 pc) with the Submillimeter Array (SMA) toward
the center of the Seyfert 2 galaxy M51.
The image shows a strong concentration at the nucleus and
weak emission from the spiral arm to the northwest.
The integrated intensity of the central component in
$^{12}$CO(J = 3 -- 2) is almost twice as high as that in
$^{12}$CO(J = 1 -- 0), indicating that the molecular gas within
an $\sim80$ pc radius of the nucleus is warm ($\gtrsim100$ K) and
dense ($\sim10^{4}$ cm$^{-3}$).
Similar intensity ratios are seen in shocked regions in our Galaxy,
suggesting that these gas properties may be related to AGN or starburst
activity.
The central component shows a linear velocity gradient
($\sim1.4$ km s$^{-1}$ pc$^{-1}$) perpendicular to the radio
continuum jet, similar to that seen in previous 
observations and interpreted as a circumnuclear
molecular disk/torus around the Seyfert 2 nucleus.
In addition, we identify a linear velocity gradient
(0.7 km s$^{-1}$ pc$^{-1}$) along the jet.
Judging from the energetics, the velocity gradient can be explained
by supernova explosions or energy and momentum transfer from the jet
to the molecular gas via interaction, which is consistent with
the high intensity ratio.
\end{abstract}


\section{INTRODUCTION}
\label{intro}

According to the unified scheme as originally suggested by
\citet{ant85}, two types (Types 1 \& 2) of Active Galactic Nuclei
(AGNs) originate from the same object (a supermassive black hole
surrounded by an optically-thick torus) that is viewed at different
angles.
This hypothesis has stimulated intensive millimeter-wave
interferometric observations to search for molecular tori toward
AGNs.
For example, images of the prototypical Seyfert 2 galaxy NGC 1068
show a $\sim100$ pc disk or torus of molecular gas around its AGN,
with a rotational axis parallel to the axis of its radio continuum
jet \citep{jac93,tac94,hel95}.
Sub-arcsecond $^{12}$CO(J = 2 -- 1) observations suggest that this
torus/disk is warped and viewed nearly edge-on at a radius of less
than 110 pc \citep{sch00}.
On the other hand, recent surveys of Seyfert galaxies suggest that
the circumnuclear molecular gas has a wide variety of distributions,
and does not always exhibit a centrally-peaked component
\citep[e.g.,][]{koh01}.
It may be that only a fraction of AGNs have molecular tori detectable
in current observations, making detailed studies of such examples
especially important.

M51 (NGC 5194), a grand-design spiral galaxy at a distance
D $\sim$ 8.4 Mpc \citep*{fel97}, has a Type 2 AGN \citep*{ho97}.
As in the case of NGC 1068, \citet{koh96} showed that the AGN of M51
is surrounded by a $\sim100$ pc disk/torus of dense molecular gas
with a rotational axis parallel to the axis of the central radio jet
\citep{for85}.
The estimated column density of the molecular gas exceeds
$10^{23}$ cm$^{-2}$, which is comparable with the amount of obscuring
material inferred from X-ray observations \citep[e.g.,][]{ter01}.
$^{13}$CO and HCN J = 1 -- 0 observations suggest that
the circumnuclear molecular gas is dense ($\sim10^{5}$ cm$^{-3}$) and
warm \citep[$\gtrsim300$ K;][]{mat98,mat99}.
These conditions suggest that the nuclear disk/torus can be better
studied in higher-J transitions, such as $^{12}$CO J = 3 -- 2
(the critical density for this transition is
 $\sim5\times10^4$ cm$^{-3}$, and the energy of the J = 3 level
 corresponds to 33 K ).

Here, we report observations of the central region of M51 with
the Submillimeter Array\footnotemark[7]
\footnotetext[7]{The Submillimeter Array is a joint project between
the Smithsonian Astrophysical Observatory and the Academia Sinica
Institute of Astronomy and Astrophysics, and is funded by
the Smithsonian Institution and the Academia Sinica.}
\citep*[SMA;][]{ho04} in the $^{12}$CO(J = 3 -- 2) line.
As anticipated, the central molecular torus/disk emits stronger in
the J = 3 -- 2 than the J = 1 -- 0, but we also detected emission
from a spiral arm.
We point out that this torus/disk exhibits a velocity gradient not
only along its major axis, but also along its minor axis parallel
to the radio jet.

\section{OBSERVATIONS}
\label{obs}

The submillimeter-wave aperture synthesis images toward the center
of M51 were obtained in the $^{12}$CO(J = 3 -- 2) line
(rest frequency = 345.796 GHz) with the SMA.
The observations were made on 2003 June 12 and 14 during the testing
and commissioning phase when a total of four 6 m antennas were
available.
The SMA correlator had a 0.96 GHz bandwidth at the time, and was
configured to have a frequency resolution of 0.8125 MHz.
We observed Mars and Uranus, respectively, for bandpass and flux
calibration, and J1310+323 every 20 minutes for amplitude and phase
calibration.
The uncertainty in the absolute flux scale is estimated to be
$\sim$ 20\%.

We calibrated the data using the OVRO software package MIR, which was
modified for the SMA.
The images were CLEANed using the NRAO software package AIPS, and
the resultant angular resolution was $3\farcs9\times1\farcs6$
(160 pc $\times$ 65 pc) at a position angle (P.A.) of $146\arcdeg$
with natural weighting.
We made channel maps with 42 channel binning, which corresponds to
a velocity resolution of 29.6 km s$^{-1}$.
The typical rms noise level of the channel maps was
70.4 mJy beam$^{-1}$, which corresponds to
T$_{\rm sys, DSB}\sim700$ K.
The half-power width of the primary beam at 345 GHz is $36\arcsec$
(1.5 kpc).
We did not detect any continuum emission at the rms noise level
attained of 18 mJy beam$^{-1}$.

\section{RESULTS}
\label{res}

Channel and integrated intensity maps in $^{12}$CO(J = 3 -- 2) as
shown in Figures~\ref{fig-channel} and \ref{fig-intmap} show a strong
peak at the nucleus at redshifted velocities \citep[with respect
to the systemic velocity of 472 km s$^{-1}$;][]{sco83,tul74}.
The highest velocity of the compact nuclear emission is
$\sim616$ km s$^{-1}$ (as can be seen also in the position-velocity
diagram of Fig.~\ref{fig-pv}), comparable with the highest velocity
seen in the redshifted wing of the $^{12}$CO(J = 3 -- 2) spectrum
taken with the JCMT \citep{mat99}.
Our maps also show weak emission from the spiral arm to
the northwest of the nucleus (hereafter, the northwestern arm) at
blueshifted velocities.
The spiral arm to the south-east (hereafter, the southeastern arm)
was not detected.
The previously published single-dish $^{12}$CO(J = 3 -- 2) maps
\citep{wie99} also show the intensity asymmetry, which is stronger
emission from the northwestern than from the southeastern arm.

The compact nuclear emission also appears at redshifted velocities
in interferometric $^{12}$CO(J = 1 -- 0) observations by
\citet{aal99} and \citet{sak99}.
Similarly, interferometric $^{12}$CO(J = 2 -- 1) observations by
\citet{sco98} also show predominantly redshifted emission,
concentrated 1'' west of the nucleus, but in addition relatively weak
blueshifted emission east of the nucleus (see their Figs.~1 and 2).
The integrated intensity of their blueshifted emission is only 18\%
of that of the redshifted emission, with a deconvolved size for
the blueshifted emission of less than $0\farcs5$.
In our map, the redshifted emission was detected at a peak integrated
intensity of $15\sigma$ at a beam size of $3\farcs9\times1\farcs6$.
Assuming the redshifted and blueshifted emission in J = 3 -- 2 has
the same intensity ratio as that in J = 2 -- 1, the blueshifted
emission is expected to have a signal-to-noise ratio of $<2\sigma$.
We rule out the possibility that the systemic velocity of
the circumnuclear molecular gas is different from that of the host
galaxy (472 km s$^{-1}$).
Interferometric HCN(J = 1 -- 0) observations by \citet{koh96} show
a compact nuclear component with comparable bright redshifted and
blueshifted emission symmetrically placed either side of the nucleus,
interpreted as an approximately edge-on rotating disk or torus.

The missing flux in our map was estimated by comparing our data
with previous single-dish observations.
The flux within the central $13\arcsec$ and $24\arcsec$ region in
our $^{12}$CO(J = 3 -- 2) map (corrected for primary beam
attenuation) is 13.8 K km s$^{-1}$ (228 Jy beam$^{-1}$ km s$^{-1}$)
and 8.8 K km s$^{-1}$ (499 Jy beam$^{-1}$ km s$^{-1}$) respectively.
The single-dish flux are 43.1$\pm$2.0 K km s$^{-1}$ within
$13\arcsec$ \citep{mat99} and 47$\pm$7 K km s$^{-1}$ within
$24\arcsec$ \citep{dum01}.
We therefore recovered only 32\% and 19\% of the single-dish flux
for the central $13\arcsec$ and $24\arcsec$ respectively, suggesting
that there exists extended $^{12}$CO(J = 3 -- 2) emission in
the central region.
Indeed, the single-dish $^{12}$CO(J = 3 -- 2) maps show very similar
image with the single-dish J = 2 -- 1 or 1 -- 0 maps, namely, even
J = 3 -- 2 emission is rich in extended structures
\citep{wie99,dum01}, which support our result.
The missing flux of $\sim70-80\%$ is, however, larger than that in
the past extragalactic $^{12}$CO(J = 1 -- 0) observations of $<50\%$
\citep[e.g.,][]{sak99}.
The possibility of calibration errors in single-dish data should be
low, because the above values are consistent with each other, and
also consistent with the independent HHT
\citep[$49\pm2$ K km s$^{-1}$;][]{mau99} and the CSO
\citep[48.3 K km s$^{-1}$ with $24''$ beam;][]{bas90} observations.
An underestimate of the absolute flux in our data is possible, but
this would suggest more extreme conditions at the central region of
M51 (see the next paragraph and section).

We compared the intensities of the nuclear molecular gas components
using our $^{12}$CO(J = 3 -- 2) data with the interferometric
J = 1 -- 0 data \citep{sak99}.
First, we convolved our J = 3 -- 2 image to the same resolution as
that in J = 1 -- 0 ($4\farcs2\times3\farcs4$).
The J = 3 -- 2 and 1 -- 0 line intensities at the nucleus averaged
over this beam size (170 pc $\times$ 140 pc) are
$91.7\pm4.8$ K km s$^{-1}$ and $49.2\pm5.2$ K km s$^{-1}$,
respectively (temperatures are in the Rayleigh-Jeans approximation,
and the uncertainties are $\pm1\sigma$).
The J = 3 -- 2 line is therefore stronger than the J = 1 -- 0 line,
with a $^{12}$CO(J = 3 -- 2)/(J = 1 -- 0) intensity ratio, $R_{31}$,
of $1.9\pm0.2$.
This line ratio may be a lower limit, as the J = 1 -- 0 data have
shorter baselines than the J = 3 -- 2 data.
To remedy this, we truncated the inner $uv$-plane of the J = 1 -- 0
data to match that in the J = 3 -- 2 data, and recomputed the line
intensities following the above method.
The resultant $R_{31}$ at the nucleus averaged over this beam size
($3\farcs9\times2\farcs6$ or 160 pc $\times$ 110 pc) is
$\sim5.2\pm1.7$, where the larger uncertainty reflects the noisier
J = 1 -- 0 image due to the inner $uv$ truncation.
In the following discussion, we adopt $R_{31}$ of $1.9\pm0.2$,
keeping in mind that this is probably a lower limit.

We made position-velocity (PV) diagrams along and perpendicular to
the axis of the radio continuum jet (P.A.\ of $\sim165\arcdeg$)
from the nucleus (Fig.~\ref{fig-pv}).
To compute the magnitude of the velocity gradient, we first searched
for the strongest peak (exceeding $2\sigma$) within a 5'' radius of
the nucleus at each velocity bin (width of 14.8 km s$^{-1}$) in
the PV diagram.
We then applied a linear fit to these peaks weighted by their
respective intensities.
In this way, we inferred a linear velocity gradient perpendicular to
the jet axis of $1.39\pm0.01$ km s$^{-1}$ pc$^{-1}$, and parallel to
the jet axis of $0.65\pm0.01$ km s$^{-1}$ pc$^{-1}$.
The sense of the velocity gradient perpendicular to the jet
(Fig.~\ref{fig-pv}a) is the same as that found in the past
interferometric observations \citep{koh96,sco98};
the magnitude of the velocity gradient of 1.39 km s$^{-1}$ pc$^{-1}$
is comparable with that estimated from the interferometric
HCN(J = 1 -- 0) observations of $1-2$ km s$^{-1}$ pc$^{-1}$
\citep{mat98}.
This velocity gradient is attributed to a nearly edge-on rotating
molecular disk or torus around the nucleus.
The velocity gradient along the jet
has not been previously discussed, but can also be seen in
the channel/velocity maps in both CO J = 1 -- 0 and 2 -- 1
(see Fig.~2 of \citealt{sco98} and Fig.~1q of \citealt{sak99}).
Based on these arguments, we believe the existence of a velocity
gradient parallel in addition to that perpendicular to the radio jet
to be real, although we note that the circumnuclear emission is
poorly resolved and hence the actual magnitude of the velocity
gradients may need to be revised when observations at higher angular
resolutions become available.

\section{DISCUSSION}
\label{dis}

The $R_{31}$ within an $\sim80$ pc radius of the nucleus of M51 is
$1.9\pm0.2$, more than twice as high as the ratios measured by
single-dish telescopes at lower spatial resolutions; line ratios
observed with the beam sizes of $\sim14\arcsec - 24\arcsec$
($\sim600-1000$ pc) are $0.5-0.8$ \citep{mat99,mau99,wie99}.
The physical properties of the circumnuclear molecular gas must
therefore be considerably different from that in the disk of
the galaxy.
We estimated the physical conditions of the circumnuclear molecular
gas using the Large-Velocity-Gradient (LVG) approximation
\citep{gol74,sco74} assuming a one-zone model.
The collision rates for CO in the temperature range $10-250$ K were
taken from \citet{flo85} and $500-2000$ K from \citet{mck82}.
The velocity gradient is estimated to be about
1.39 km s$^{-1}$ pc$^{-1}$ (Sect.~\ref{res}).
This value is consistent with the gradient caused by the internal
turbulence \citep[$\sim2$ km s$^{-1}$ pc$^{-1}$; for a more detailed
discussion, see][]{mat98}.
Using the `standard' relative abundance
[$^{12}$CO]/[H$_{2}$] $=5\times10^{-5}$, we find the H$_{2}$ number
density $n_{\rm H_{2}}\gtrsim10^{4}$ cm$^{-3}$ and the kinetic
temperature $T_{\rm k}\gtrsim500\pm150$ K, with the uncertainties
reflecting the uncertainty in $R_{31}$.
Even for an order of magnitude lower relative abundance or an order
of magnitude higher velocity gradient, this model gives
$n_{\rm H_{2}}\gtrsim10^{4}$ cm$^{-3}$ and
$T_{\rm k}\gtrsim95\pm15$ K.
We therefore conclude that the circumnuclear molecular gas is denser
and warmer than the molecular gas in the galactic disk.
This is consistent with the results of the interferometric $^{13}$CO
and HCN J = 1 -- 0 observations, which also suggest
dense ($n_{\rm H_{2}}\sim10^{5}$ cm$^{-3}$) and warm
($T_{\rm k}\gtrsim300$ K) molecular gas based on LVG calculations
\citep{mat98,mat99}.

\begin{figure}
\epsscale{0.85}
\plotone{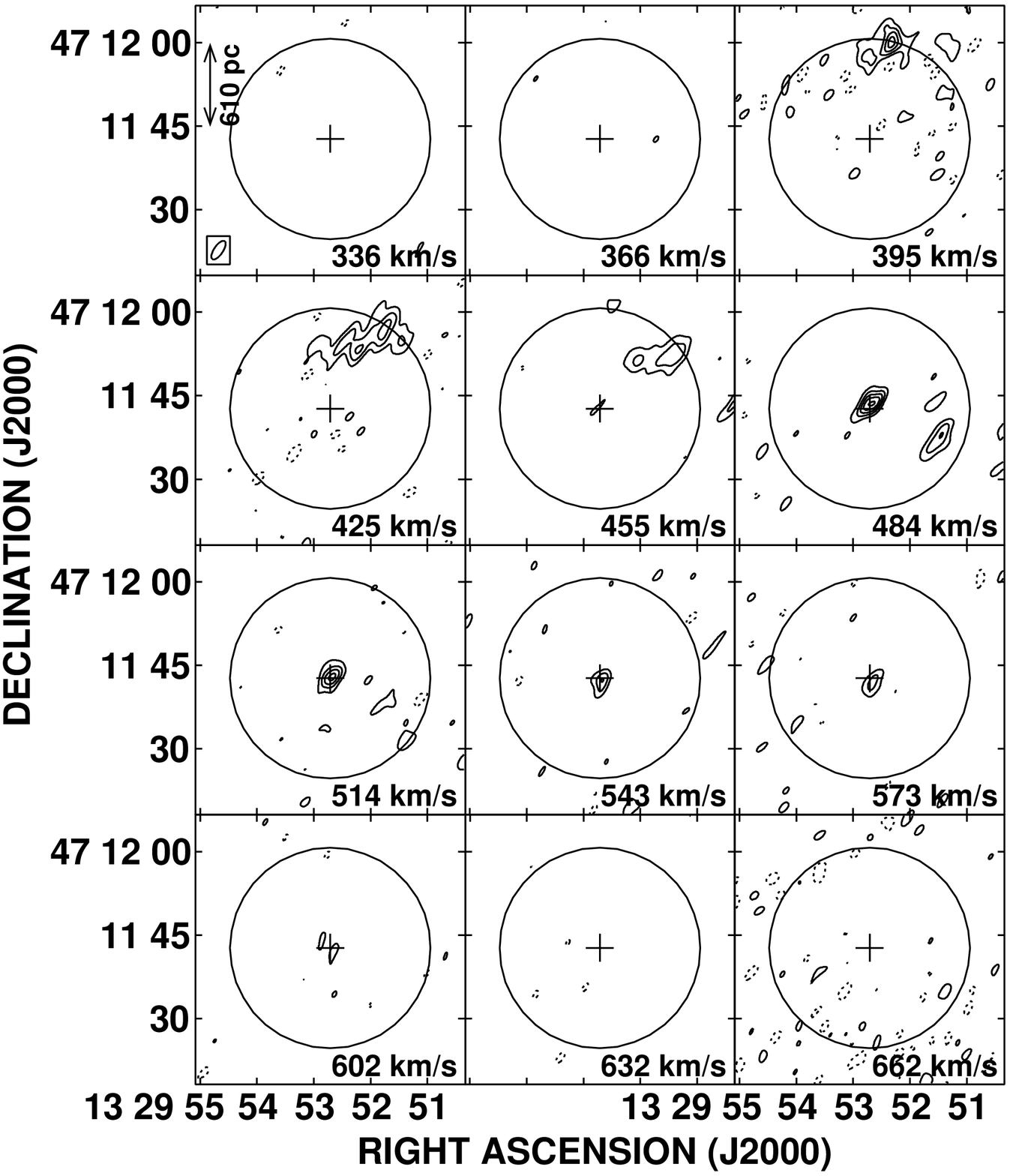}
\caption{\small
	Channel maps of the $^{12}$CO(J = 3 -- 2) line of the central
	region of M51.
	The synthesized beam ($3\farcs9\times1\farcs6$ or
	160 pc $\times$ 65 pc) is shown at the lower-left corner of
	the first channel map.
	LSR velocities are shown at the bottom-right corner of each map.
	The systemic velocity is 472 km s$^{-1}$ \citep{sco83,tul74}.
	The cross and the $36\arcsec$ (1.5 kpc) diameter circle in each
	map are the galactic nucleus determined from the 6 cm radio
	continuum peak \citep{for85} and the half-power width of
	the primary beam.
	The contour levels are $-6, -3, 3, 6, 9, 12,$ and $15\sigma$,
	where $1\sigma$ = 70.4 mJy beam$^{-1}$ (= 0.115 K).
	Primary beam correction is not applied.
\label{fig-channel}}
\end{figure}

\begin{figure}
\plotone{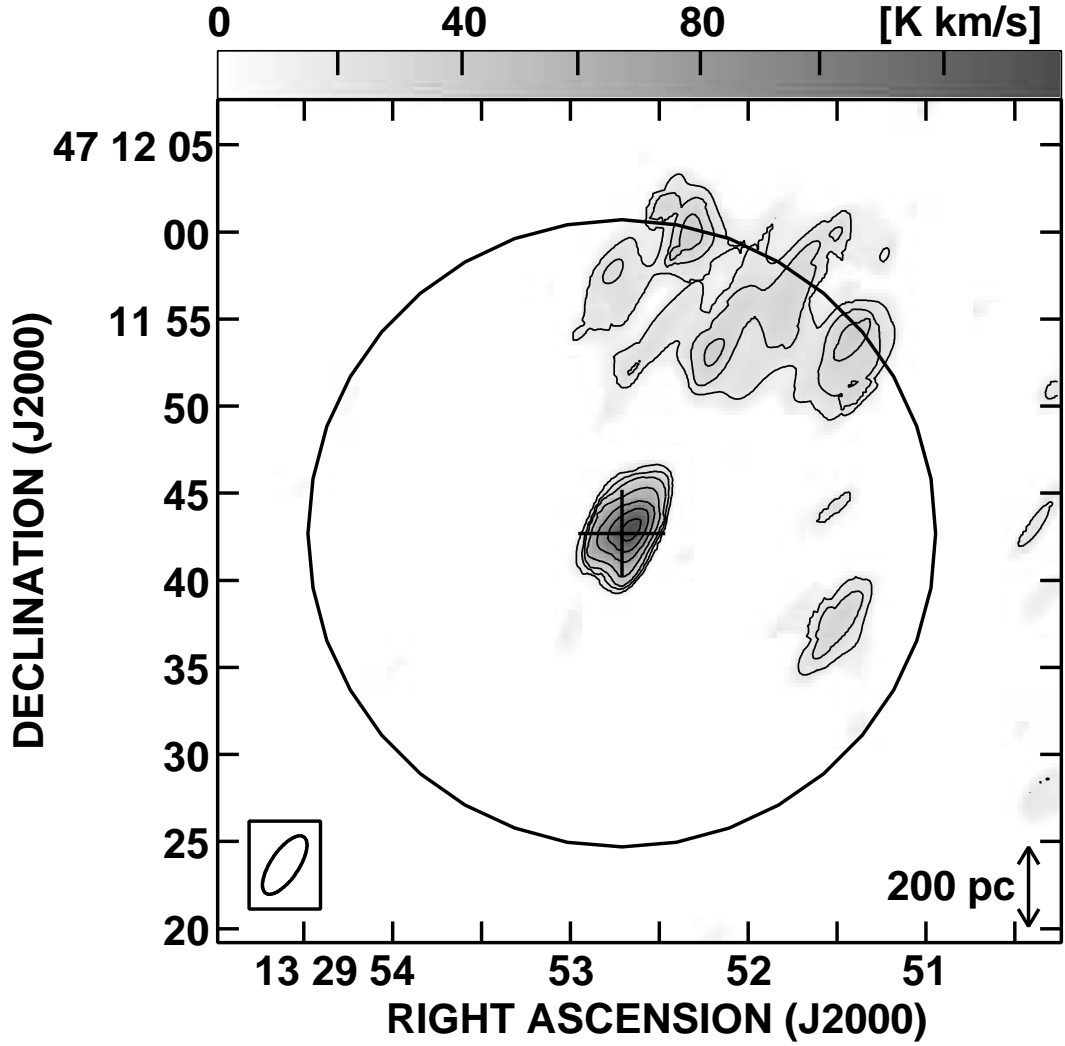}
\caption{
	Integrated intensity map of the $^{12}$CO(J = 3 -- 2) line.
	The gray scale and the synthesized beam are shown at the top
	and the bottom-left corner, respectively.
	The cross and the circle are the same as in Fig~\ref{fig-channel}.
	The contour levels are $2, 3, 4, 6, 9, 12$ and $15\sigma$,
	where $1\sigma$ = 5.89 Jy beam$^{-1}$ km s$^{-1}$
	(= 9.63 K km s$^{-1}$).
	Primary beam correction is not applied.
\label{fig-intmap}}
\end{figure}

\begin{figure}
\plotone{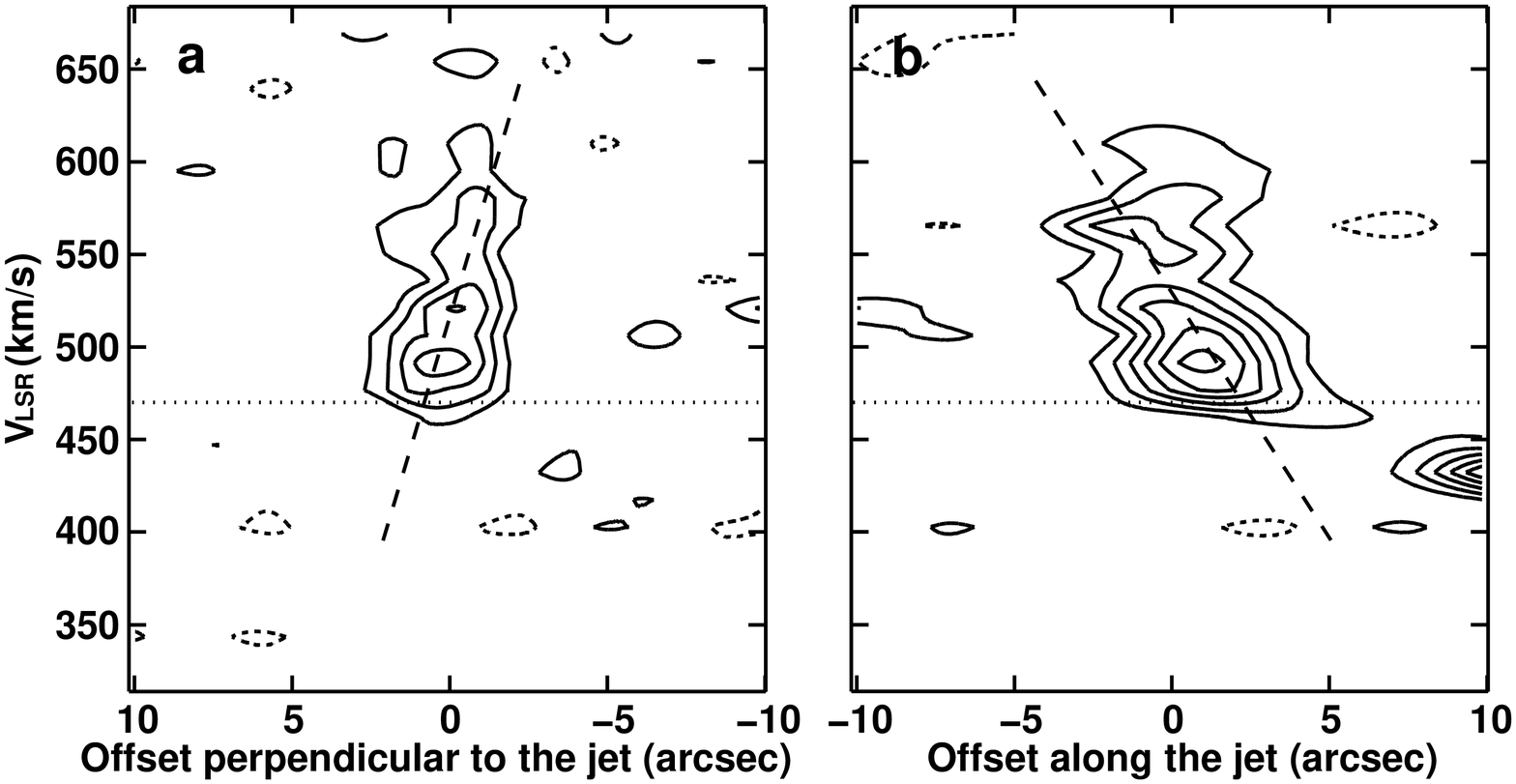}
\caption{
	Position-velocity diagrams (a) along and (b) perpendicular
	to the radio jet (P.A. $\sim165\arcdeg$).
	The velocity resolution is 14.8 km s$^{-1}$.
	Positions are the offsets from the nucleus.
	The contour levels are $-2, 2, 4, 6, 8, 10$ and $12\sigma$,
	where $1\sigma$ = 76.5 mJy beam$^{-1}$ for (a) and
	47.8 mJy beam$^{-1}$ for (b).
	Dotted horizontal lines are the systemic velocity.
	Dashed lines show the results of intensity-weighted
	linear fitting.
\label{fig-pv}}
\end{figure}

There are several possible reasons for the rather extreme conditions
of the molecular gas.
One is interaction with the radio jet.
Molecular outflows from star forming regions in our Galaxy show
$R_{31}$ greater than unity \citep[e.g.,][]{ric85,hir01}, similar to
that in the center of M51.
The strongest $^{12}$CO(J = 3 -- 2) emission in our channel maps
comes from $1\arcsec$ north-west of the nucleus
(484 km s$^{-1}$ in Fig~\ref{fig-channel}),
and this also is a region with high [\ion{N}{2}]/H$\alpha$ intensity
ratio ($\sim4$).
The high [\ion{N}{2}]/H$\alpha$ ratio suggests that relativistic
particle heating is dominant \citep{cec88}, which suggests
an interaction between the jet and molecular gas.
Recent high resolution H$_{2}$O maser observations reveal that
the masers are located close to the nucleus, and are possibly
associated with the jet \citep{hag01}.
The velocity range of the H$_{2}$O masers of $538-592$ km s$^{-1}$
is similar to that of the redshifted nuclear emission in our data
(see Fig.~\ref{fig-channel} and \ref{fig-pv}), which also supports
the idea that the molecular gas is interacting with or entrained by
the jet.
Shock/compression by supernova (SN) explosions is another
possibility, since shocked molecular gas around SN remnants in our
Galaxy also shows high intensity ratios \citep[e.g.,][]{ari99}.

The linear velocity gradient {\it along} the jet axis is
an interesting phenomenon, and it is worth considering what may
cause this gradient.
We first calculate the mass, momentum, and kinetic energy of
the molecular gas within the velocity range of
515 -- 615 km s$^{-1}$, which is away from the systemic velocity.
The CO-to-H$_{2}$ conversion factor can be estimated using
the LVG approximation mentioned above, and the result is
$\sim1\times10^{20}$ cm$^{-2}$ (K km s$^{-1}$)$^{-1}$.
The mass is calculated using this conversion factor and
the intensities of molecular gas.
In each channel map within the velocity range, the velocity of
the molecular gas is defined as the velocity difference from
the systemic velocity, and we assume that the molecular gas is
moving along the line of sight.
Thus the estimated values are lower limits.
Adding all channel map information, we estimated the mass, momentum,
and kinetic energy as $1\times10^{6}$ M$_{\odot}$,
$2\times10^{46}$ g cm s$^{-1}$, and $8\times10^{52}$ erg,
respectively.

Using these values, we discuss the possible causes of the velocity
gradient along the jet axis.
One is that the molecular gas away from the systemic velocity is
entrained by the jet, namely, relativistic particles.
The minimum energy of the jet is $6.9\times10^{51}$ erg
\citep{cra92}, and the momentum is $2\times10^{41}$ g cm s$^{-1}$,
if we assume that the velocity of the relativistic particles is 90\%
of the speed of light.
\citet{cra92} estimated the energy of the jet assuming the total
energy of the jet plasma is 100 times larger than that of electrons.
This number can be larger, but it should be smaller than the mass
ratio of proton and electron of about 1800.
The actual geometry of the jet and the molecular gas is unclear, but
should be similar if molecular gas is entrained by the jet.
Therefore the energy of the jet may be similar to that of the molecular
gas, but the momentum is several orders of magnitude different.
This problem is also seen in the study of young stellar objects, and
still under discussion \citep[e.g.,][]{ric00}.
Another possibility is that since the momentum and energy can be
transferred from one to another by interaction.
If those of the molecular gas along the jet are continuously input by
the jet via interaction, the momentum and energy differences between
the molecular gas and the jet can be explained.
Indeed, the high [\ion{N}{2}]/H$\alpha$ ratio (an evidence of
relativistic heating) and the H$_{2}$O masers (an evidence of shocked
molecular gas) are observed, as mentioned above.
These observations support the idea that the energy and momentum
of the molecular gas along the jet are transferred from the radio
jet via interaction.

There are several possible explanations to create the velocity
gradient along the jet without considering interaction with the radio
jet.
Supernova (SN) explosions can be one of the possible causes, based on
the similarity of $R_{31}$ as mentioned above.
However, $10^{2}-10^{4}$ SNe are needed to explain the energy of
the molecular gas, if we assume that one SN releases $10^{51}$ erg
and that the energy transfer efficiency is at most 20\% \citep{mcc87}.
Alternatively, the velocity gradient along the jet may be due to
a warped disk around the AGN \citep[e.g.,][]{sch00}.
In this case, not all the gas is on a plane of a molecular gas disk.
Thus, even if most of the gas show velocity gradient perpendicular to
the jet, some of the gas show velocity gradient along the jet.
Future SMA observations with higher spatial resolution will provide
more detailed information about the molecular gas around the AGN.

\acknowledgements

We thank P.T.P.\ Ho, J.M.\ Moran, and M.J.\ Cai for valuable comments.
We also thank all the past and present SMA staffs for designing,
constructing, and supporting the SMA.

\clearpage

\end{document}